\documentclass[a4paper]{jpconf}
\usepackage{graphicx}
\begin{document}
\title{A large millikelvin platform at Fermilab for quantum computing applications}

\author{Matthew Hollister, Ram Dhuley and Grzegorz Tatkowski}

\address{Fermi National Accelerator Laboratory, P.O. Box 500, Batavia, IL 60510, USA}

\ead{mhollist@fnal.gov}

\begin{abstract}
The need for larger mK cooling platforms is being driven by the desire to host ever growing numbers of cryogenic qubits in quantum computing platforms. As part of the Superconducting Quantum Materials and Systems Center at Fermilab funded through the Department of Energy under the National Quantum Initiative, we are developing a cryogenic platform capable of reaching millikelvin temperatures in an experimental volume of 2~meters diameter by approximately 1.5~meters in height. The platform is intended to host a three-dimensional qubit architecture based on superconducting radiofrequency accelerator cavity technologies. This paper describes the baseline design of the platform, along with the expected key performance parameters.
\end{abstract}

\section{Introduction}
\label{sect:introduction}

The Superconducting Quantum Materials and Systems Center (SQMS) at Fermilab aims to understand the physics and materials origin of coherence limiting factors in quantum devices. An extremely promising approach to increasing the coherence time of superconducting qubits is to utilize high quality factor superconducting cavities, a technological area that has seen extensive development for use in superconducting radiofrequency (SRF) particle accelerators, combined with two-dimensional devices. In such a configuration, the two-dimensional superconducting circuit will couple to multiple degrees of freedom in the three-dimensional cavity to produce multiple interconnected quantum states that can each function as a discrete qubit. The operation of niobium-based SRF cavities at millikelvin temperatures has previously been demonstrated to exhibit long lifetimes in the quantum regime~\cite{romanenko20}, although there are challenges in cooling large superconducting objects to such low temperatures due to the large heat capacity through the transition temperature and the poor thermal conductivity of the superconductor below the transition. The SRF cavity qubit architecture results in a relatively large object that must be cooled to millikelvin temperature in order to suppress photon noise in the cavity, while the use of multiple such devices drives to a large experimental volume combined with a high cooling capacity in order to cool the cavities to base temperature.

To address the basic cryogenic requirements of the SRF cavity architecture, SQMS plans to construct a large cryogenic platform capable of supporting multiple cavities and/or conventional two-dimensional qubit devices in a large volume cooled to approximately 20~mK with large cooling capacity. The current largest dilution refrigerators available on the commercial market have millikelvin volumes in the 500--1000~mm diameter range and cooling capacities of approximately 40~$\mu$W at 20~mK. The SQMS Center targets a platform with a 20~mK volume 2~meters in diameter with a cooling capacity 200--300~$\mu$W at 20~mK.

In addition to exploring the coherence time of qubit devices, a second area of interest that is a use case for the large millikelvin platform are ``dark photon'' searches in which an emitter cavity is used to produce a signal that can be detected in a second receiver cavity that is well-isolated from the emitter (see, for example~\cite{parker13}). The technique has been successfully demonstrated with both the emitter and receiver cavities at 1.4~K in a modified magnet vertical test stand at Fermilab, but in order to increase the sensitivity of the receiver cavity lower temperature operation is required. To this end, the platform is also designed to host an emitter cavity operating at a temperature $\sim$2~K along with several receiver cavities operating at 20~mK.

This paper will discuss the main design features of the millikelvin platform, along with the current status and expected timeline for completion of construction.


\section{Challenges of scaling up millikelvin platforms}
\label{sect:challenges}

Large~\cite{ligi16} and high-cooling power dilution refrigerators~\cite{frossati87} have been previously been constructed, but this projects calls for a combination of both features. There are two major limiting factors in the construction of larger fridges, which are the surface area of the millikelvin heat exchangers required to overcome the kaptiza boundary resistance of high 3He circulation rates, and the available cooling capacity at the higher temperatures stages of the refrigerators, particularly in the 2--4~K range.

\subsection{Overcoming limitations of heat exchangers}
The cooling capacity of the dilution process is primarily a function of the rate of circulation of 3He in the circuit. However, the circulation rate cannot be increased arbitrarily since the increased flow rate requires increasing heat transfer between the inflow and outflowing 3He streams in the heat exchangers. Particularly at the lowest temperatures near the mixing chamber, the required area for heat exchange quickly becomes unreasonably large. Rather than attempt to design larger surface area heat exchangers, the approach taken by this project is to use multiple, parallel dilution refrigerators housed in the same cryostat with the common temperature stages tied together by the cold plates. The use of multiple, smaller heat exchanger stacks has the additional advantage that existing commercial design can be utilized, and the exchanger stacks can be added gradually to upgrade the system capacity with time. The platform is designed to mount up to ten dilution stages, each with an independent room temperature pumping and gas handling system.

\begin{figure}[t]
\begin{center}
\includegraphics[width=0.95\textwidth,angle=0]{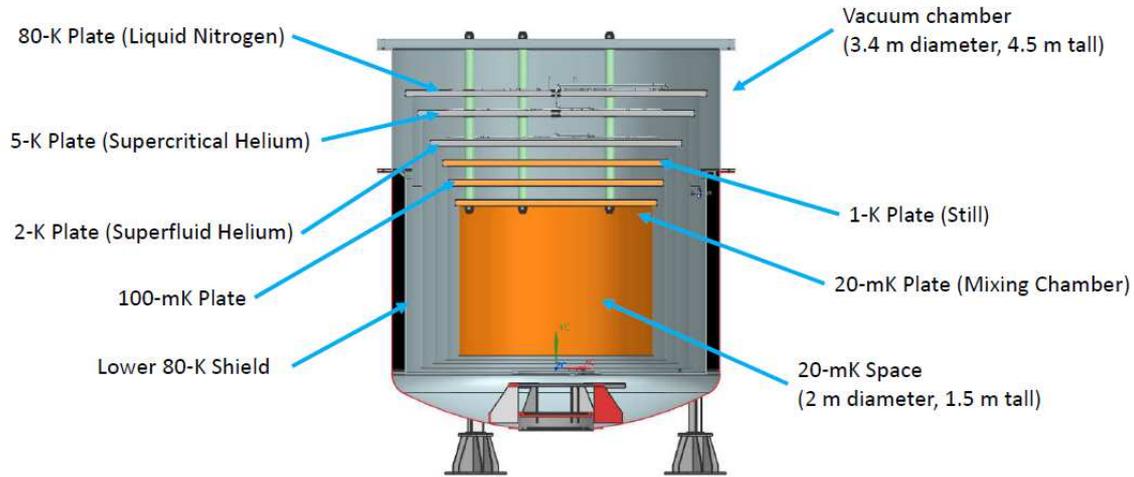}
\end{center}
\caption{\label{fig:cryostat}CAD image of the cryostat design shown in cross-section with the major features of the design indicated. The dimensions of the outer vacuum shell are 3.4 meters diameter and 4 meters high. For details and discussion, see text.}
\end{figure}

\subsection{Cooling power limitations at intermediate temperature stages}
\begin{figure}[p]
\begin{center}
\includegraphics[width=1.0\textwidth,angle=0]{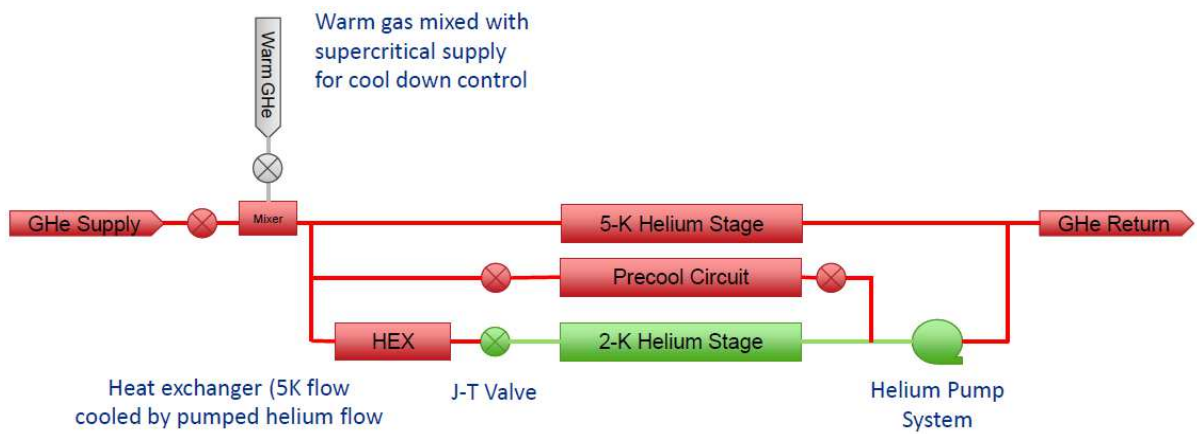}
\end{center}
\caption{\label{fig:5KBlock}Block diagram schematic of the 5-K and 2-K cooling systems. For detailed discussion, see text.}
\end{figure}

The load at the intermediate stages of the dilution refrigerators come from a combination of conductive loads from supports and wiring, and the load from cooling and condensing the circulating helium. While older designs of dilution refrigerator utilized liquid helium to provide cooling at temperatures around 4~K as a precooling stage and to supply a pumped 4He pot (the ``1-K Pot'') as a precooling stage, the consumption of liquid helium imposes a practical limit on the heat load that can be absorbed. More modern designs have made use of closed-cycle cryomechanical coolers which, while having the advantage of no cryogen consumption, are extremely limited in their ability to take up heat at temperatures of 4-K. This is addressed in the SQMS platform by using a cooling configuration that is probably more familiar to particle accelerator designs in which a cryoplant is used in refrigeration mode to supply supercritical helium at a temperature of 5~K to the cryostat. The flow is divided, with one channel directly cooling a 5-K cold plate and thermal shield, with the other stream being subcooled in a small heat exchanger and expanded through a Joule-Thomson (J-T) valve to produce a two-phase mixture of liquid and gaseous helium. The pressure of the downstream side of the J-T valve is reduced to approximately 30~mbar (equvialent to a temperature of 2~K) by a room temperature pumping system to provide a powerful condensation stage analogous to the 1-K Pot. This configuration is shown schematically in Fig.~\ref{fig:5KBlock}, with additional discussion in the next section. The cryoplant and pump system are closed-cycle, eliminating the issue of helium consumption. Cooling at the warmer cryogenic stages is provided by an open-cycle liquid nitrogen flow.


\section{Design of the SQMS mK platform}
\label{sect:design}

\begin{figure}[p]
\begin{center}
\includegraphics[width=1.0\textwidth,angle=0]{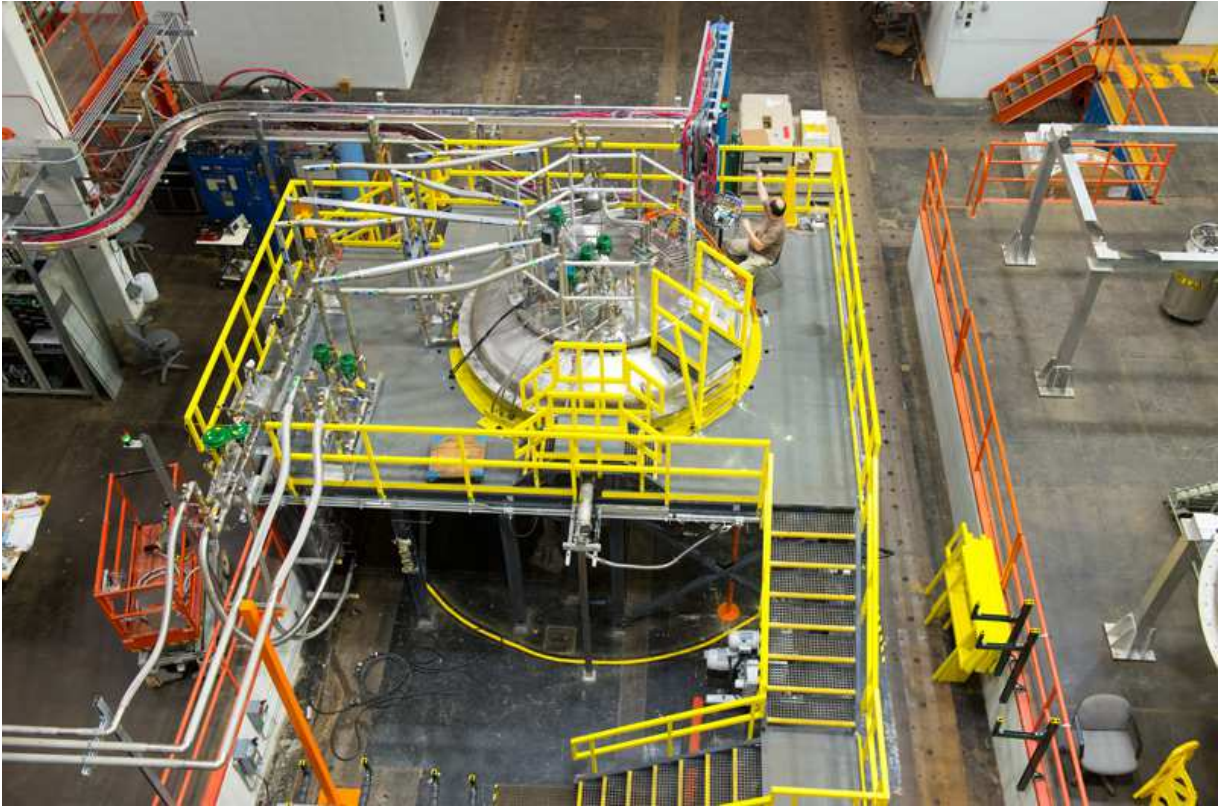}
\end{center}
\caption{\label{fig:OVC}Photograph of the existing installation of the vacuum vessel, including the associated work platform and cryogenic infrastructure. The vacuum vessel was formally used as a liquid helium test stand for solenoids for the Mu2e project. Picture credit: Fermilab}
\end{figure}

The general layout of the millikelvin platform in cross section is shown in Fig.~\ref{fig:cryostat}. The platform consists of a flat upper head from which the internal assemblies are suspended. This assembly is installed in a stainless steel vacuum vessel which also supports a thermal shield cooled by a liquid nitrogen circuit. The lower vessel and thermal shield were originally used as part of an existing test cryostat, while the flat head and internal assembly are a new construction. The existing cryostat is shown in Fig.~\ref{fig:OVC} with a dished vessel head installed. The structure suspended from the flat head consists of six cold plates at temperatures of 80~K cooled by the liquid nitrogen supply, 5~K and 2~K cooled by the helium flow, and at 1~K, 100~mK and 20~mK cooled by the stages of the dilution refrigerator themselves. The 5~K, 2~K, 1~K (still) plates support shields for thermal radiation control, while the 20~mK plate mounts a shield for control of stray light in the experimental space. The major dimensions and mechanical parameters for the cryostat are summarized in Table~\ref{mechanicalparameters}.

\begin{table}[p]
\caption{\label{mechanicalparameters}Dimensions and construction materials of the major cryostat components. For the shield components, ``Thickness'' refers to the dominant wall thickness of the shell and ``Diameter'' is the inside diameter of the shield cylinder.}
\begin{center}
\begin{tabular}{lcccc}
\br
Component             & Diameter & Height & Thickness & Material \\
\mr
Vacuum Vessel         & 3.4~m    & 3.7~m & 10~mm & SS304 \\
Thermal Shield        & 3.0~m    & 2.8~m & 17~mm & SS304 \\
Helium Shield         & 2.7~m    & 2.6~m & 5~mm & AL6061 \\
2-K Shield            & 2.5~m    & 2.2~m & 5~mm & AL6061 \\
Still Shield          & 2.2~m    & 2.0~m & 5~mm & AL6061 \\
Mixing Chamber Shield & 2.0~m    & 1.5~m & 5~mm & OFHC \\
Vacuum Plate          & 3.6~m    & -     & 90~mm & SS304 \\
Thermal Shield Plate  & 3.0~m    & -     & 50~mm & AL6061 \\
Helium Plate          & 2.8~m    & -     & 50~mm & AL6061 \\
Superfluid Plate      & 2.5~m    & -     & 50~mm & AL6061 \\
Still Plate           & 2.3~m    & -     & 25~mm & OFHC \\
IC Plate              & 2.2~m    & -     & 25~mm & OFHC \\
Mixing Chamber Plate  & 2.0~m    & -     & 25~mm & OFHC \\
\br
\end{tabular}
\end{center}
\end{table}

The various thermal stages are suspended from the flat vacuum head using vertical tie rods. To reduce the stress in the rods due to the radial contraction of the large cooled plates, the rods are not rigidly attached but instead connect via hinged joints that allow for radial movement of the rods while the cold plates remain horizontal. The exact configuration of the materials and dimensions of the tie rods is currently being optimized, but will likely consist of a mixture of stainless steel and G-10CR epoxy composite.

For the purposes of maintenance and payload installation, the vacuum vessel head and suspended structures are removed from the vacuum shell using an overhead gantry crane and lowered through the roof of an adjacent clean workroom. The lower thermal shield remains installed in the vacuum shell during this operation.

\subsection{Warm cryogenic stages}
\label{sect:warmcryo}

Cooling at the upper stages of the cryostat cold mass is provided by flow of liquid nitrogen at approximately 80~K and two stages cooled by the flow of supercritical helium. The liquid nitrogen system is fed from a 20000 gallon storage tank outside the building and is supplied to the cryostat via a vacuum jacketed supply line and a custom manifold consisting of pneumatic valving and bayonets. The supply line for the nitrogen system, along with the supply and return lines for the helium supply, can be see in the lower left of the cryostat imagine in Fig.~\ref{fig:OVC}. Three liquid nitrogen supply bayonets are provided, with one feeding the lower thermal shield, one the Thermal Shield cold plate in the main cold mass, and the third supplying an open-cycle liquid nitrogen dewar that is used to cool the cold traps in the 3He gas system to remove air impurities. Both the stainless steel cooling lines on the cold plate and on the lower thermal shield meander around the structures in order to provide distributed cooling. Copper clamps are attached the cooling line at intervals, and then bolted to the cold plate to provide the thermal connection.

\begin{figure}[p]
\begin{center}
\includegraphics[width=0.9\textwidth,angle=0]{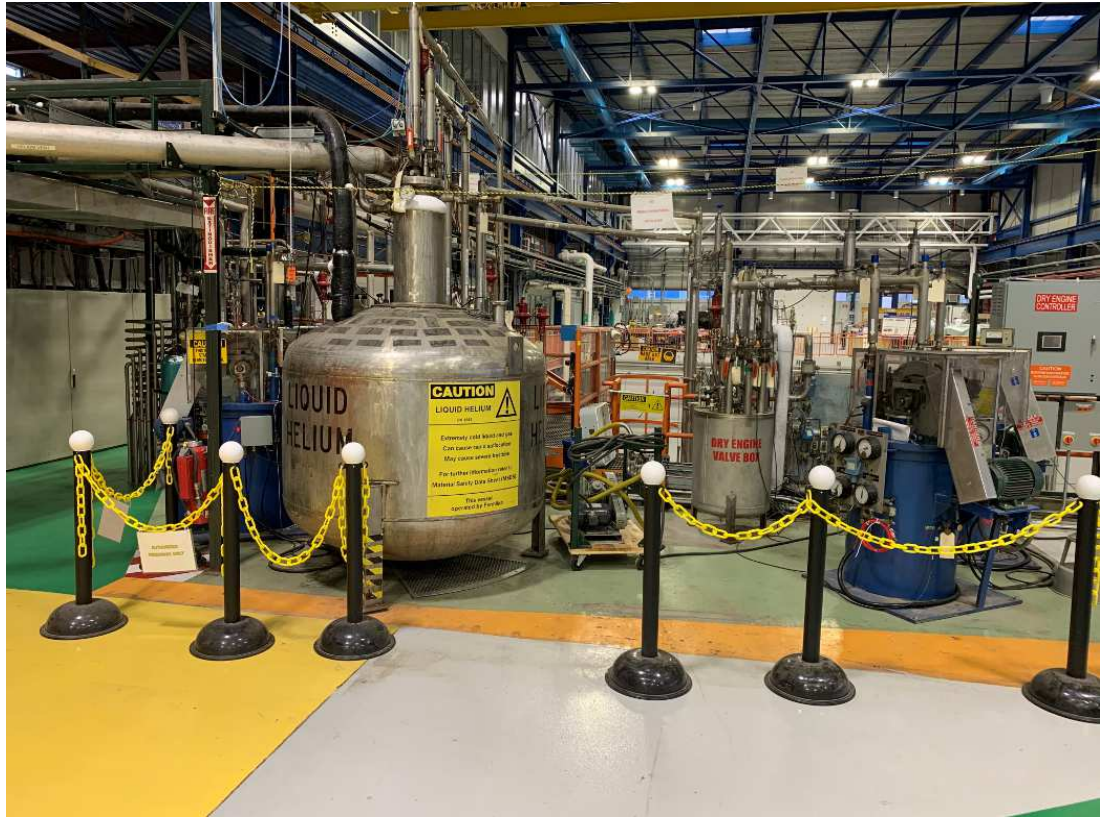}
\end{center}
\caption{\label{fig:cryoplant}Photograph of the helium cryoplant, including the 2000 liter liquid helium dewar on the left. The plant is rated at a refrigeration capacity of 625~W at 4.5~K or 4.2 g/s of liquid helium.}
\end{figure}

For cooling at helium temperatures, supercritical helium is supplied from a cryogenic plant via a vacuum jacketed transfer line, paired with a corresponding return line. The cryogenic plant, which can be seen in Fig.~\ref{fig:cryoplant} consists of a compression system and reciprocating expanders that can act either as a helium liquifier or a refrigerator supplying supercritical gas, or a combination of the two modes.~\cite{rode83} For the supercritical supply, the system is capable of providing up to 625~W of cooling at 4.5~K. The cold helium gas is supplied to the cryostat at a pressure of approximately 50~psig and on entering the platform cryostat, is divided into one or more flow paths, as shown schematically in Fig.~\ref{fig:cryoplant}. The primary 5-K flow path is similar to the liquid nitrogen piping system for 80~K cooling on the upper stage, and typically carries approximately 20~\% of the flow while the remaining flow is diverted through either the precooling circuit during the initial cool down of the cryostat, or through a subcooling heat exchanger and Joule-Thomson valve in normal operation. The precooling system directs flow to each of the sub-2-K plates in series with the return path connected to the header of the superfluid pumping system discussed below. During normal operation, the precooling system must be evacuated of helium in order to avoid any superfluid flow leaks in the precooling system that would increase the thermal load on the lower temperature stages of the system. To allow control of the cool down process, the supercritical helium entering the cryostat passes through a static mixer with a controlled flow of warm helium gas from the cryoplant suction header provided to regulate the temperature of the gas flow in both the precool circuit and the main 5-K circuit.

Cooling at the 2-K stage of the cryostat is provided by a superfluid helium system. The flow diverted to the system during normal operation is subcooled in a small counterflow heat exchanger with the outflowing cold vapor, and is then expanded through a Joule-Thomson valve to produce a two-phase mixture. The pressure of the mixture is lowered by pumping with a room temperature pumping system consisting of a Kinney vacuum booster with a liquid ring pump. The pump line exiting the cryostat is a 2~inch diameter stainless steel pipe, which transitions to a 4~inch diameter vacuum jacketed pipe outside the cryostat for connection to the pumping system, approximately 200~feet away from the platform vacuum vessel. The heat exchanger and pumping system is sized for a cooling capacity of approximately 50~W. As noted previously, the superfluid cooling system is used to provide an additional thermal intercept for conductive loads in the cryostat, but more importantly provides the condensation stage of the 3He circulating in the dilution circuits. This accounts for only a small fraction of the available cooling power at the superfluid stage (see summary in Table~\ref{thermalmodelresults}), with the remaining capacity available for cooling a transmitter cavity when the platform is used for Dark SRF experiments.

\begin{figure}[ht]
\begin{center}
\includegraphics[width=1.0\textwidth,angle=0]{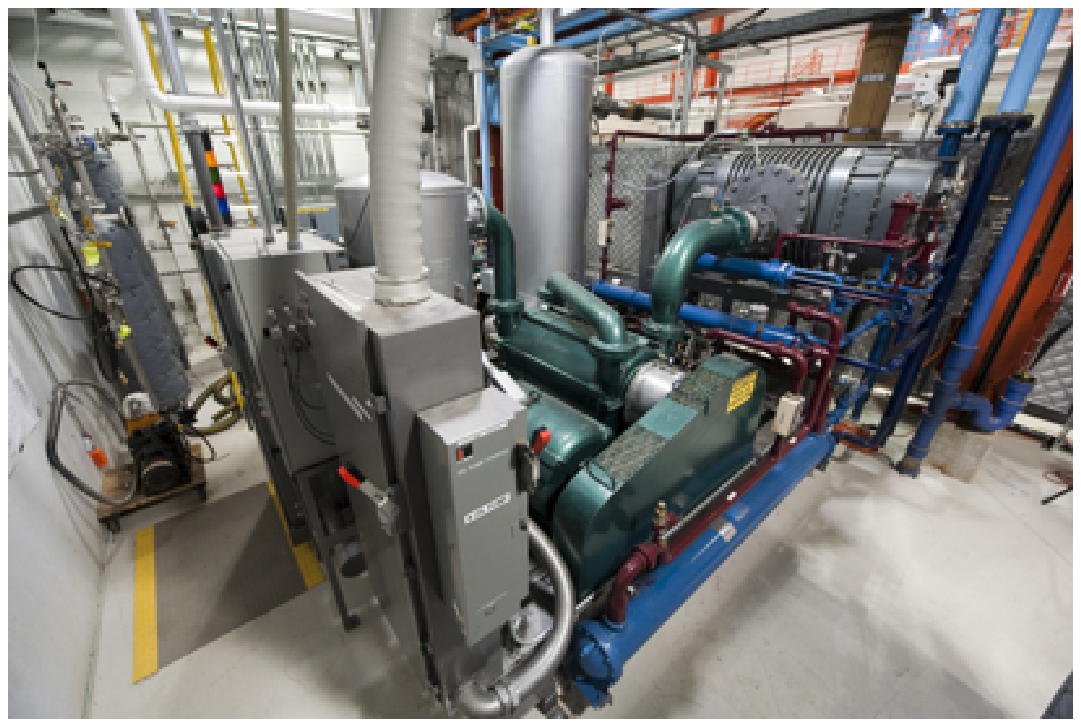}
\end{center}
\caption{\label{fig:2KPump}Example of a pumping system for the superfluid helium system, consisting of a Kinney vacuum booster and liquid ring pump.}
\end{figure}

Both the outlet of 2-K pumping system and the return line of the 5-K cooling system connect to low pressure room temperature storage tanks, with a total capacity of 40000 gallons. The collected gas is compressed using a Mycom screw compressor skid, with the high pressure gas being supplied back to the reciprocating engines.

\subsection{Sub-1~K cryogenic stages}
\label{sect:coldcryo}

Cooling of the three coldest stages of the experiment is provided by the dilution unit heat exchanger stacks. As mentioned previously, the platform utilizes commercially-constructed dilution heat exchangers, with capacity for up to ten independent 3He circuits to be installed. It is expected that initially only a subset of the dilution stages will be installed, allowing the platform to operate albeit at reduced cooling capacity, with later expansion to the full system. The specifics of the heat exchanger stacks are being finalized, but based on the currently available commercial dilution units it is likely that each stack will be able to provide 30~$\mu$W at 20~mK at the mixing chamber and approximately 10 mW at 800~mK at the still layer, depending on the room temperature pumping system. The mixing chamber of each dilution unit is bolted directly to the 20~mK plate, while thermal connections at the intermediate cold plate and the still layer are made via flexible copper ropes to minimize thermal stresses in the stacks due to contraction of the large cold plates.

The still side of the dilution units are each pumped by an independent gas handling system. Each system has provision for either 1 or 2 large turbo pumps, each turbo system being backed with an Edwards GXS-series dry screw pump. Unlike modern cryogen-free dilution refrigerators that typically require a diaphragm compressor or similar to increase the condensing pressure, the high working outlet pressure of the dry screw pumps combined with the use of a 2~K condensation stage allows this system to operate with the primary pump only. 3He on the return side of the pump system passes through a charcoal cold trap cooled by liquid nitrogen in order to remove any air or other impurities.

On entering the cryostat, the inlet line first passes through a filter packed with metal sinter that is cooled by the 80~K stage of the cyrostat. This filter both removes further impurities from the flow while also acting as the first heat exchanger to cool the incoming gas. The flow then enters a spiral heat exchanger at the 5-K stage of the cryostat consisting of approximately 1~meter of 2~mm diameter cupronickel tube wrapped onto a copper plug of 25~mm diameter that is bolted to the 5-K cold plate. A second, similar spiral heat exchanger is used to the 2-K cold plate in which the 3He is further cooled and condensed, before flowing on to the still heat exchanger via an impedance.

It is estimated that each dilution circuit will be charged with 150 STP liters of 4He and 50 STP liters of 3He. The two helium isotopes are stored in separate room temperature tanks for each dilution circuit to allow for easy tuning of the mixture.

\subsection{Thermal modelling}
\label{sect:thermal}

Detailed modelling of the expected thermal loads at each stage of the cryostat was conducted during the design process in order to specify the required capacities for each part of the system. The model included thermal radiation, conductive loads, and loads from the circulation of helium mixture in the dilution circuits. The results of the model are summarized in Table~\ref{thermalmodelresults} in comparison to the available capacity at each stage. To allow for unexpected loads and for possible future expansion, a margin of 60~\% or higher has been maintained at each stage. Of particular note is the contribution of the RF readout wiring, which is assumed to consist of a flexible strip line arcitecture using conductive traces on Kapton films, with 1000 readout lines included in the model. If a more conventional semirigid coaxial cable arrangement were used, it would be expected that the contribution of this term would be higher but still easily within the allowable margin of the design. In this platform, the cold HEMT amplifiers are assumed to be at the 2~K stage rather than at the 4~K stage as is commonly seen in other systems. The power dissipation of these cold amplifiers is included in the RF Wiring term at this stage in Table~\ref{thermalmodelresults}.

\begin{figure}[p]
\begin{center}
\includegraphics[width=0.8\textwidth,angle=0]{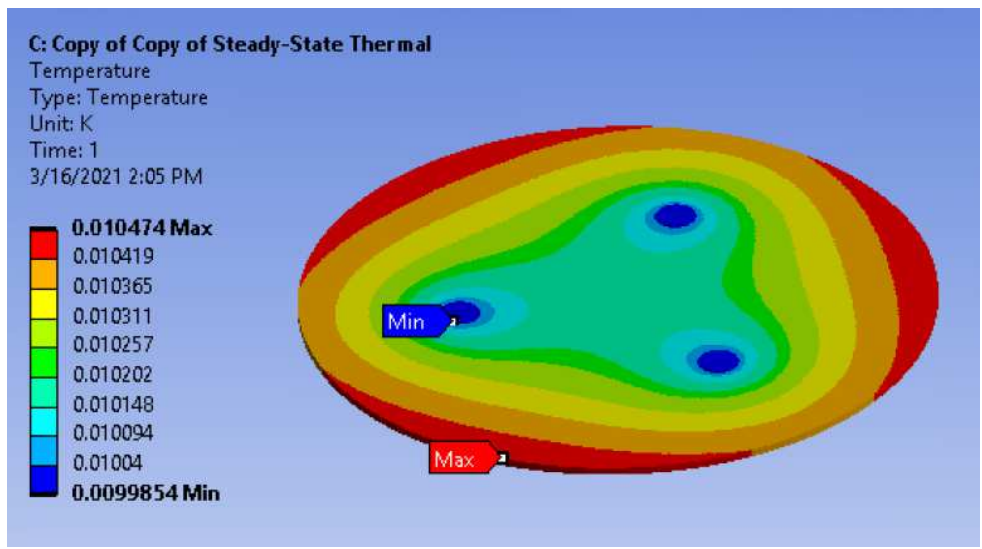}
\end{center}
\caption{\label{fig:fea}Sample results of FEA analysis for the 20-mK cold plate. The plate is composed of copper with an assumed residual resistivity ratio of 50, 50~mm thick and 2~m diameter. Three points represent the connections between the plate and three mixing chambers. With the mixing chamber contacts fixed at 10~mK and a uniform 50~$\mu$W applied to the outer edge of the plate, the maximum temperature is 10.5~mK.}
\end{figure}

In addition to the static thermal calculations, the temperature distributions in the large cold plates and thermal shields under power loading was explored using finite-element modelling. These models were used to inform the material selections and dimensions at each stage. An example of these simulation results is shown in Fig.~\ref{fig:fea} for the mixing chamber cold plate at 10~mK. This is a copper plate with thickness of 50~mm showing a temperature gradient of order 0.5~mK with 50~$\mu$W applied uniformly to the outer edge. The use of copper at the lower temperature stages is important due to the discrete point cooling of the dilution units, while at the higher temperature stages with distributed cooling lines carry helium and nitrogen, aluminum alloy can be used with acceptable temperature gradients. Similarly, aluminum alloy shields can used at all stages with the exception of the mixing chamber shield.

\begin{table}[p]
\caption{\label{thermalmodelresults}Summary of thermal modelling results listed by cryostat stage and major loads.}
\begin{center}
\begin{tabular}{lcccccc}
\br
Load & Thermal  & Helium & Superfluid & Still & Intermediate & Mixing  \\
     & Shield   &        & Helium     &       & Cold Plate   & Chamber \\
\mr
\emph{Nominal}     &      &     &     &     &        &      \\
\emph{Temperature} & 80 K & 5 K & 2 K & 1 K & 100 mK & 20 mK \\
\mr
Radiation & 23 W & 1.8 W & 23 $\mu$W & 0.49 $\mu$W & - & - \\
RF Wiring & 7 W & 1 W & 1.11 W & 753 $\mu$W & 407 $\mu$W & 2.7 $\mu$W \\
DC Wiring & 2.2 W & 2.1 W & 29 mW & 0.31 $\mu$W & 0.05 $\mu$W & - \\
Supports & 33.7 W & 4.4 W & 9 mW & 1 mW & 198 $\mu$W & 2.2 $\mu$W \\
Precool Line & & & & & & \\
Conduction & - & - & 370 $\mu$W & 43 $\mu$W & 8.5 $\mu$W & 0.1 $\mu$W \\
Helium-3 Inlet & & & & & & \\
Conduction & 1.7 W & 0.2 W & 0.4 mW & 25 $\mu$W & - & - \\
Mixture Cooldown & 91.5 W & 31.2 W & 2.24 W & - & - & - \\
Still Pumpline & & & & & & \\
Conduction & 74.5 W & 2.7 W & 2 mW & 0.49 $\mu$W & - & - \\
2-K Pumpline & & & & & & \\
Conduction & 15.8 W & 1.8 W & 23 $\mu$W & - & - & - \\
\mr
\textbf{Total Load} & 250 W & 65 W & 3.4 W & 4 mW & 614 $\mu$W & 5 $\mu$W \\
\mr
\textbf{Available} &     &       &      &        &      &            \\
\textbf{Capacity} & 9kW & 200 W & 50 W & 100 mW & 1 mW & 300 $\mu$W \\
\br
\end{tabular}
\end{center}
\end{table}

\section{Current status and timeline}
The detailed final design process is nearing completion as of summer 2021 with major procurements expected to start during 2021. The initial construction phase for the upper cryogenic stages is planned to complete late in 2022, allowing operation of the platform as a large 2-K cryostat for Dark SRF experiment development. Upgrade of the platform to millikelvin operation is planned for 2023 with full operation as a quantum computing platform by the end of 2023.

\newpage

\section{Summary}
This paper has discussed the design for a large millikelvin cryogenic platform suitable for use as a quantum computing platform and as a platform of Dark sector photon searches using superconducting radiofrequency cavities. The major design details have been discussed, along with the expected heat loads and thermal performance metrics associated with the system. The platform will address issues associated with scaling up millikelvin cryogenic systems by utilizing existing cryogenic infrastructure for the supply of helium and liquid nitrogen, along with integrating multiple small dilution circuits into a single large cryostat to provide a large, high-cooling power cryostat. The design phase is nearing completion, with procurements expected to begin in 2022 with initial construction and start of operations during 2023.

\begin{ack}
This material is based upon work supported by the U.S. Department of Energy, Office of Science, National Quantum Information Science Research Centers, Superconducting Quantum Materials and Systems Center (SQMS) under the contract No. DE-AC02-07CH11359.
\end{ack}

\end{document}